**Manuscript Title:** Objective estimation of sensory thresholds based on neurophysiological parameters

**Abbreviated Title:** Sensory threshold estimation

Achim Schilling[1], Richard Gerum[2], Patrick Krauss[1], Claus Metzner[2], Konstantin Tziridis[1], and Holger Schulze[1]

[1] Experimental Otolaryngology, ENT-Hospital, Head and Neck Surgery, Friedrich-Alexander University Erlangen-Nürnberg (FAU), Germany

[2] Department of Physics, Center for Medical Physics and Technology, Biophysics Group, Friedrich-Alexander University Erlangen-Nürnberg (FAU), Germany

# Abstract

Reliable determination of sensory thresholds is the holy grail of signal detection theory. However, there exists no gold standard for the estimation of thresholds based on neurophysiological parameters, although a reliable estimation method is crucial for both scientific investigations and clinical diagnosis. Whenever it is impossible to communicate with the subjects, as in studies with animals or neonatales, thresholds have to be derived from neural recordings. In such cases when the threshold is estimated based on neuronal measures, the standard approach is still the subjective setting of the threshold to the value where at least a "clear" neuronal signal is detectable. These measures are highly subjective, strongly depend on the noise, and fluctuate due to the low signal-to-noise ratio near the threshold. Here we show a novel method to reliably estimate physiological thresholds based on neurophysiological parameters. Using surrogate data, we demonstrate that fitting the responses to different stimulus intensities with a hard sigmoid function, in combination with subsampling, provides a robust threshold value as well as an accurate uncertainty estimate. This method has no systematic dependence on the noise and does not even require samples in the full dynamic range of the sensory system. It is universally applicable to all types of sensory systems, ranging from somatosensory stimulus processing in the cortex to auditory processing in the brain stem.

# Introduction

Objective, reliable and reproducible estimation of sensory thresholds is a fundamental problem in neuroscience as well as clinical diagnostics. For example, hearing thresholds must be determined as objectively and precisely as possible in patients with hearing loss, especially in those who cannot report their hearing, e.g. babies [1], to provide them with the optimal type of hearing aid and to adjust the operating parameters of the device. Similarly, the determination of thresholds from physiological measurements in animals is a challenging task [2].

A fundamental problem with common methods for threshold estimation is that these usually use responses just above threshold, and trivially these responses are small and thereby strongly affected by noise. In other words, common methods for threshold estimation are based on data with low signal-to-noise ratio (S/N), where responses may be hard to detect or cannot be detected at all (miss) so that thresholds are often assessed too large. In particular, automated methods that use a statistical criterion to define the threshold, e.g. the signal amplitude several standard deviations above background noise [1,3,4], are prone to such threshold overestimation. On the other hand, also false positive ratings (false alarms) may occur, especially if data are evaluated subjectively by human observers. A striking example for such severe uncertainty is the evaluation of auditory brainstem responses (ABR) by clinical professionals, where threshold estimates have been demonstrated to differ by up to 60 dB between evaluators [5]. Finally, as responses are repeatedly measured and then averaged, the S/N strongly depends on the number of repetitions, which introduces a further source of error to the threshold estimation. Obviously, a fully automated method for threshold estimation which is robust against low S/N would be preferable to guarantee objectivity and reproducibility. This challenging problem has also been tackled by the implementation of Machine learning algorithms such as support vector machines [6]. However, machine learning approaches generate black-box systems with complex internal decision criteria that are not comprehensive to the users [7]. In addition, they require huge data sets for training and hence are neither feasible nor accepted for medical diagnostic purposes [8].

We here introduce a method for threshold estimation that solves the problems mentioned above and that can be applied to any type of neuronal data for threshold estimation, i.e. to any data of response strength as a function of stimulus intensity. Our method is robust against low S/N, as no measurements of responses close to threshold have to be included into the analysis.

We used simulated data to evaluate the objectivity, reproducibility and robustness of the method. In addition, we demonstrate the method's feasibility with real ABR data and with cortical neuronal responses (local field potentials (LFP) and single neuron spiking responses) from different sensory modalities (auditory and somatosensory) in an animal model.

We show that the fitting of stimulus-response functions to neural responses can be significantly improved by taking into account the spontaneous neural background activity under non-stimulus conditions. Furthermore, we demonstrate that any threshold criterion based on the fit function should not depend on the spontaneous activity amplitude as such criteria lead to a monotonic decrease (divergence to $-\infty$) of the determined threshold with decreasing noise amplitude or increasing number of measurement repetitions. Thus, a generalized hard sigmoid function was chosen to fit the data, where the lower knee of the function defines the sensory threshold, resulting in a minimum set of three free parameters and a completely objectified method for threshold estimation based on neurophysiological data.

# Results

## Artificial test data

The major problem for the verification of any method for threshold estimation lies in the fact that net stimulus response amplitudes are unknown due to variable amounts of noise. To test our new approach for reliable threshold estimation we generate an artificial test data set (Fig. 1; for details cf. Online Methods). This allows for the verification of estimated thresholds by comparing them to the thresholds on which the artificial data is based. To this end, a stimulus-response function was defined for the ideal case of no measurement noise (Fig. 1a). This stimulus response function is then translated to artificial neuronal responses (raw signal) approximated by a 1000 Hz sine wave (Fig. 1b). The addition of Gaussian distributed background noise (Fig. 1c) simulating real measurement noise results in realistic raw data (Fig. 1d).

For each stimulus intensity $x$, we generate 200 independent samples of such artificial raw neural recordings. As in a real data evaluation, these recordings are then averaged to reduce the noise (Fig. 1e). Based on the resulting average signal, the root-mean-square (RMS) amplitude is computed. When plotted as a function of the stimulus intensity $x$, we obtain a re-constructed stimulus-response function (Fig. 1f), which in general has an altered sigmoidal shape $f(x) \neq f_0(x) + \sigma$. This re-constructed function $f(x)$ is then used to test our new method of threshold estimation.

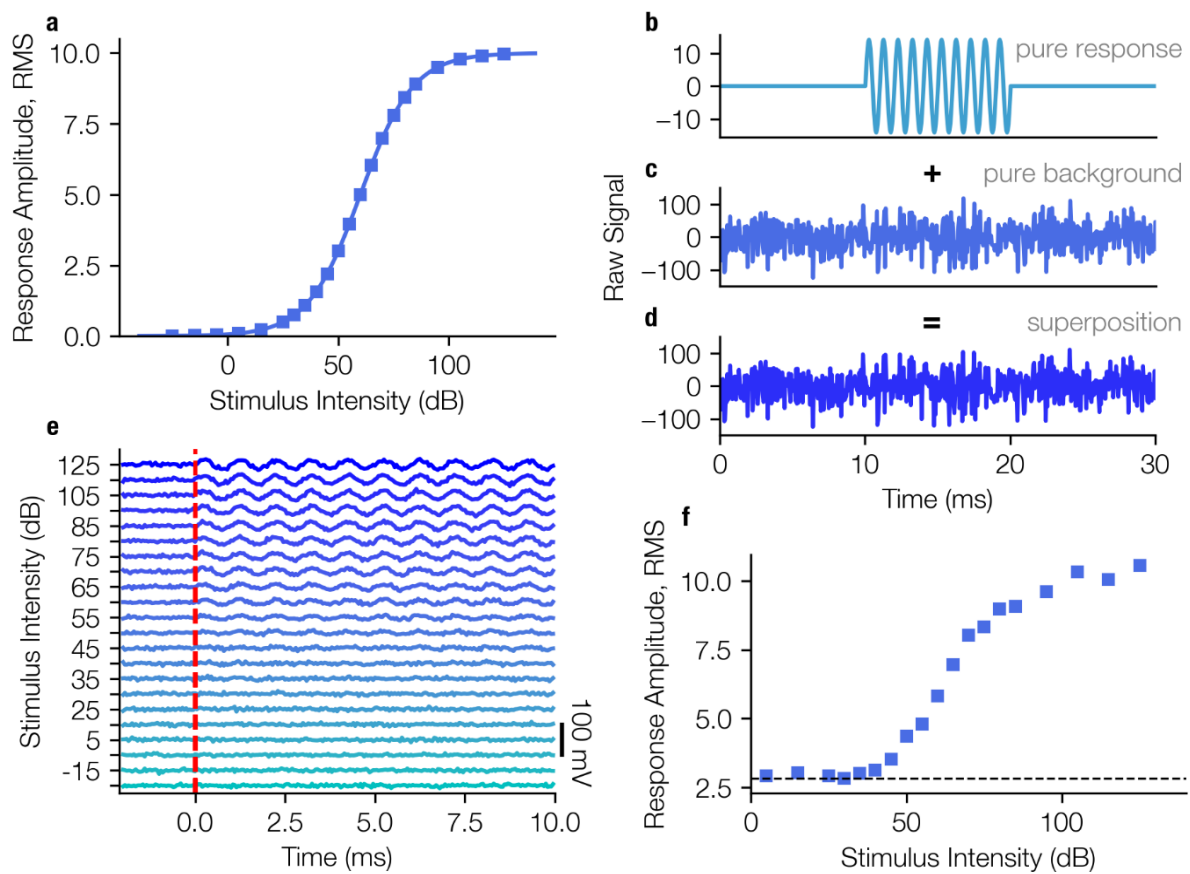

**Fig. 1 | Generation of artificial test data sets.**
**a**, Template amplitude response function used to generate the artificial field potential. **b**, The stimulus response is simulated as a sine-wave of frequency 1000 Hz and a duration of 10 ms (impulse response to short stimulus). **c**, The measurement noise is modelled as Gaussian white noise with an RMS amplitude 4 times higher than the maximal simulated response. **d**, The artificial signal is a superposition of the simulated response and the simulated background noise. **e**, Simulated neuronal signal (e.g. ABR waves superposed with background noise), each averaging 200 single trials (**d**). **f**, Response intensity function for the simulated data set in (**e**). The RMS of the background alone (dashed line) is 40/sqrt(200) being approximately 2.8.

## Threshold criteria

To estimate sensory thresholds the standard procedure is to measure the response amplitude as a function of stimulus intensities. The resulting stimulus response function typically follows a sigmoid shape, cf. [9,10] and can be fitted using a generalized logistic function $f_0$ with an offset for the background noise:

$$f_0(x) = \frac{a}{1 + e^{-\left(\frac{x-b}{c}\right)}} \quad \text{(equ. 1)}$$

$$f(x) = f_0(x) + \sigma \quad \text{(equ. 2a)}$$

$$f(x) = \sqrt{f_0(x)^2 + \sigma^2} \quad \text{(equ. 2b)}$$

where *a* refers to the maximum response, *b*, the location of the inflection point, *c*, an indirect measure for the slope at the inflection point, and σ the noise level.

Depending on the experiment, the noise either is added directly to the response (equ. 2a; e.g., for spike rates), or the response is a RMS value where the noise is added according to equation (equ. 2b, cf. Fig. 2a; e.g., for RMS values of field potentials; cf. Supplements S2 for the derivation). If the noise would also be treated as an additive term in the case of a RMS value, the signal and noise would not be decoupled correctly and the obtained thresholds would be noise dependent (see Supplements S1).

In principle all four parameters could be fitted [9,11], but fitting the noise level from the response function results in highly unstable threshold estimates (cf. Supplements S4). Therefore, we estimate the background noise level by analyzing the response without stimulation and hold the background noise level σ constant for the fitting procedure. To extract a threshold from the fitted stimulus response functions, a threshold criterion has to be defined.

There exist multiple approaches how to define a threshold criterion for a given sigmoidal function. However, defining a reliable threshold criterion is anything but trivial. In many studies threshold criteria are used that depend on the background noise, e.g. the 2σ-criterion, where the threshold is set to the point where the function exceeds two times the standard deviation of the noise, i.e. two times the RMS of the noise [12]. Another common approach is to define the threshold as a constant fraction *p* of the dynamic range *a*, e.g. the 5%-criterion, where the threshold is set to the point where the sigmoid function exceeds 0.05*a* [9]. As we will show in the following, both approaches to define a threshold criterion have fundamental drawbacks.

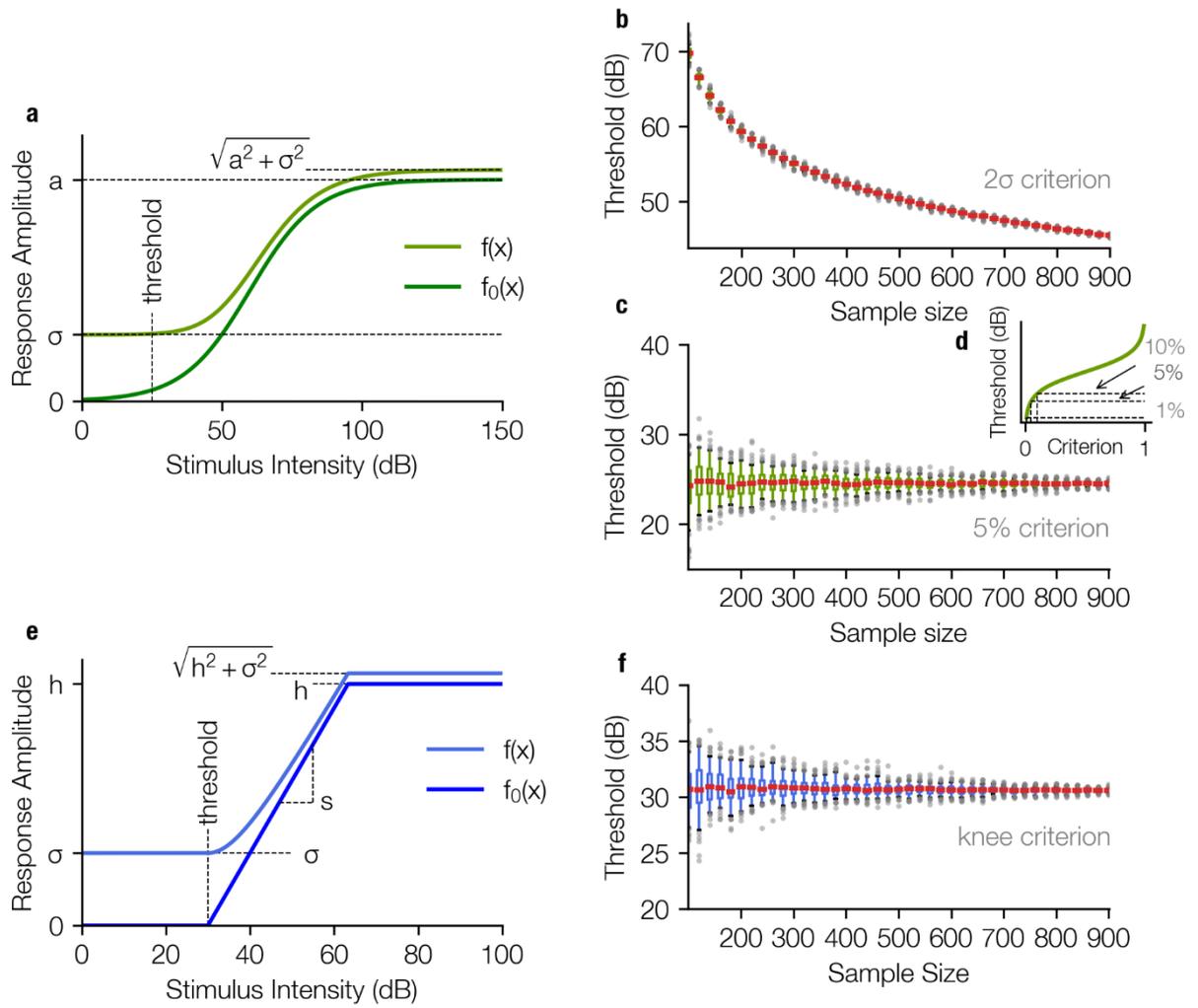

**Fig. 2 | The fit function based on a generalized logistic function.**
**a**, The stimulus response function without background noise is approximated using a generalized logistic function (dark green). The fit function (light green) is the square root of the sum of the squared "pure" neuronal signal and the squared noise amplitude (including neural noise and measurement noise, for the derivation cf. Supplements S2). **b**, The estimated threshold for the 2σ-criterion as a function of the number of applied data samples decreases monotonically with rising sample size and diverges to $-\infty$ for an infinite number of samples. **c**, The estimated threshold for the 5% criterion is independent of the sample size, only the uncertainty decreases (estimated by subsampling, cf. Methods). **d**, Different choices of the threshold parameter p for this criterion result in different threshold estimates. **e**, To get rid of the arbitrary parameter p, the sigmoid shape of the stimulus response function can be approximated with a hard sigmoid function (dark blue), where the threshold is defined as the lower knee. In analogy to the logistic function, for fitting the responses the function is superimposed with a noise offset sigma. **f**, Estimated threshold for the knee criterion as a function of the sample size. The small constant offset compared to the analysis in (**c**) is caused by the arbitrary parameter p used for the 5% criterion. The analysis in (**b**,**c**,**f**) was performed by stepwise subsampling with increasing sample size N, with 100 subsamples for each size.

The threshold based on the 2σ-criterion can be expressed as a function of the fit parameters *a*, *b*, *c* and the constant parameter σ (equ. 3).

$$f(t_\sigma) = 2\sigma$$

$$t_\sigma = -c \cdot \ln\left(\frac{a}{\sqrt{3}\,\sigma} - 1\right) + b. \qquad \text{(equ. 3)}$$

Thresholds based on the 5% criterion (*p*=0.05) can be calculated from the fit parameters *c* and *b* of the generalized logistic function:

$$f_0(t_{p=5\%}) = p \cdot a$$

$$t_{p=5\%} = -c \cdot \ln(p^{-1} - 1) + b. \qquad (\text{equ. 4})$$

The parameter *a* cancels out when transforming the equations, but still has an indirect influence on the threshold estimate as it influences the other parameters of the fit.

For both criteria, we analyze how the sample size (mimicking number of measurement repetitions), and thus the effective background noise, influences the obtained threshold values.

For a systematic analysis we use the artificial data set as described above (cf. Fig. 1), and evaluate the threshold for different sample sizes (Fig. 2b,c,f). As the background noise is Gaussian distributed and the data is averaged across all samples, the effective noise amplitude σ scales indirectly proportional to the square root of the sample size N:

$$\sigma = \frac{1}{\sqrt{N}} std(X). \qquad (\text{equ. 5})$$

Hence, for a measurement with lower background noise the number of samples across which has to be averaged to get a satisfying signal to noise ratio can be reduced.

As mathematically the effective noise level decreases with the number of samples, it is trivial to see that the 2σ-criterion, being directly noise dependent, results in a decrease of the threshold value with increasing number of samples (Fig. 2b). The fit parameters *a*, *b*, *c* approach a constant value for increasing number of samples and σ approaches 0, the threshold estimate $t_\sigma$ diverges to $-\infty$ (cf. Fig. 2b), i.e. for an infinite number of samples the threshold becomes infinitesimal.

In contrast to the 2σ-criterion, for the 5% criterion the threshold is set to the stimulus intensity where $f_0(x)$ exceeds $p \cdot a$ (p=5%), i.e. the threshold is set to the value where the net amplitude response function (with no background noise) exceeds the fraction *p* of the dynamic range (cf. equ. 4). This modified procedure leads to highly reproducible threshold estimates that are independent of the number of measurement repetitions (Fig. 2c).

The median threshold for increasing sample size converges to the threshold obtained from the net response function without noise (cf. Fig 1a). The 5% criterion overcomes the limitation of a systematic dependency of the estimated threshold on the number of measurement repetitions. Furthermore, p=5% is an arbitrary parameter on which the threshold depends and the choice of p can significantly influence the value obtained for the threshold (cf. Fig 2d).

This fitting approach has a second major shortcoming: The generalized logistic function is inadequate for the analysis of most real data as the function is symmetric around the inflection point and missing supporting points in the saturation range lead to unstable fitting of the sigmoid function, as we will show in detail below. These supporting points are often missing in measured ABR responses, as the response of bigger clusters of neurons often saturate for very high stimulus intensities [9] (exemplarily shown in Supplements S5).

## Parameter reduction using a hard sigmoid based fit function

To overcome these problems we chose a generalized hard sigmoid function – often used for artificial intelligence approaches, cf. [13] – instead of the generalized logistic function for fitting ($f_0(x)$), thereby eliminating the arbitrary parameter p and decoupling the lower and upper end of the dynamic range.

$$f_0(x) = \begin{cases} 0, & x < t \\ s \cdot (x - t), & t \leq x \cap s \cdot (x - t) < h \\ h, & s \cdot (x - t) \geq h \end{cases} \quad \text{(equ. 6)}$$

(*t*: lower knee = sensory threshold, *h*: saturation value, *s*: slope, cf. Fig. 2e)

The lower knee of this function can be defined as the sensory threshold in analogy to the procedure used for the logistic function. When noise is added, this knee is smoothed according to equ. 2b (Fig. 2e, for derivation of the smoothening of the lower knee cf. Supplements S3).

When this approach is applied to the data set shown in Fig. 1, in analogy to the approach described before (cf. Fig. 2c, for sigmoid function), the determined threshold is independent of the number of measurement repetitions (cf. Fig. 2f).

Taken together, this procedure further reduces fit parameters as the arbitrary parameter *p* is eliminated. Additionally, the novel fit function is more robust against missing data (supporting points) in the saturation range, as will be discussed in detail in the following section.

## Effect of removal of data supporting points

Naturally, when attempting to measure sensory or behavioral thresholds, the actual threshold for a given stimulus as well as the dynamic range are both unknown. Depending on the chosen intensity range of presented stimuli, response amplitudes may lie (as preferred) within the dynamic range, but stimulus intensities may also lie below threshold or above saturation levels. If too few data points are available to sample the dynamic range of the sensory response, standard fitting procedures often fail to yield meaningful results. Therefore, to further validate the robustness of our new approach we test the effect of removal of data points on the estimated threshold. We use the same simulated data set with the logistic function as underlying stimulus response function (cf. Fig.1) and then stepwise reduce the number of supporting data points.

The determined threshold for both fit functions are independent of deletion of the subthreshold supporting points (Fig. 3a,b,c). As σ can easily be calculated from the raw data measurement under the non-stimulus condition (which is mandatory) and is used in our approach as a fixed value for fitting, subthreshold data points are redundant. The shape of the curve can even be estimated if approximately half of the dynamic range supporting points are deleted (cf. Fig. 3c). Consequently, data points close to the threshold and hence with low signal-to-noise ratio do not have to be measured anymore.

Likewise, a stepwise removal of supporting points within the saturation range has little effect on the estimated threshold for the hard sigmoid fit (cf. Fig. 3e,f). In contrast to the logistic function fit, which is less robust (cf. Fig. 3d,f). This is particularly important as very high stimulus intensities for practical reasons often cannot be presented without causing potential damage to the sensory system.

Taken together, our approach to use a fitting procedure with a hard sigmoid function provides high robustness against deletion of supporting points in subthreshold and threshold as well as saturation range, reduction of the number of applied measurement repetitions (sample size), and does not depend on an arbitrary threshold parameter *p*.

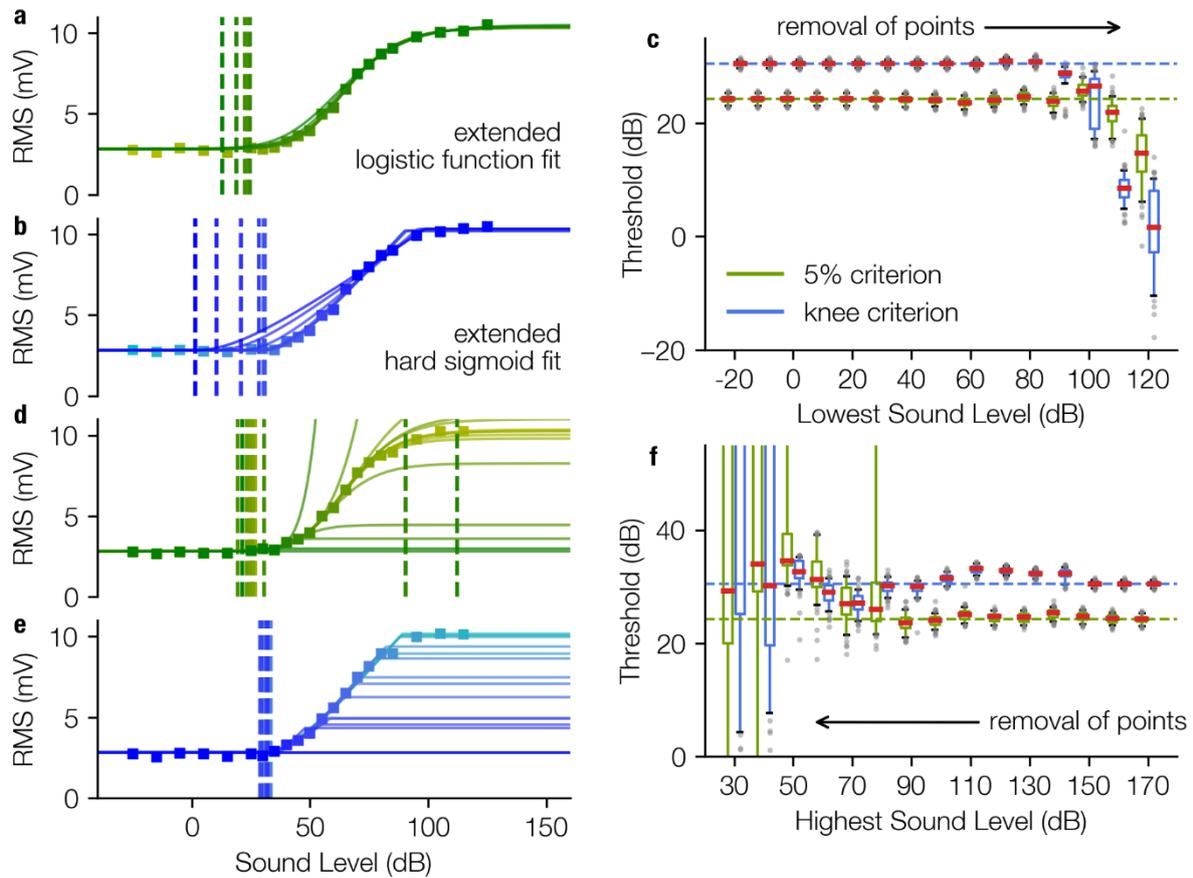

**Fig. 3 | Effect of removal of supporting points.**
**a**,**b**, Stepwise removal of the supporting points starting at the sub-threshold range for the logistic function fit with the 5%-criterion (**a**) and the hard sigmoid fit with the knee criterion (**b**). **c**, The determined threshold is very robust against deletion of supporting points near the saturation range where the S/N ratio is poor. **d**,**e**, Stepwise removal of supporting points near the saturation range (high stimulus intensities) for the logistic function fit (**d**) and the hard sigmoid fit (**e**). **f**, The hard sigmoid fit is very robust against missing supporting points near the saturation range. Note that even though the used artificial data is generated using an underlying generalized logistic function the hard sigmoid fitting procedure is more stable against missing supporting points near the saturation range. This fact is important e.g. for far field potential measurements as the dynamic range of such measurements typically is very large due to the different thresholds of the involved neurons are distributed over a wide range, cf. [9]

## Application of the method to different types of neurosensory data

So far we have developed and tested our new approach for threshold estimation based on artificial data. In the following section, we demonstrate that this approach is universally applicable and that the underlying principles can be applied to a number of different neurophysiological parameters. We performed stimulus response measurements in different brain regions of Mongolian gerbils (brainstem, cortex), different sensory modalities (auditory, somatosensory system) and different measures of evoked neurophysiological activity (far field potentials, spiking activity).

## a) Auditory brainstem responses (ABR)

ABR waves in response to pure tone stimuli of 6 ms duration with onset and offset ramps (2ms) and four different stimulus frequencies (1kHz, 2kHz, 4kHz, 8kHz) were measured.

The response amplitude is quantified by the calculation of the RMS over the signal in the 10 ms time interval after stimulus onset (Fig. 4a,b), resulting in a typical stimulus response relationship (Fig. 4c,d). This relationship can be well described by the hard sigmoid fit, yielding threshold estimates with the knee criterion.

To estimate the confidence of these obtained sensory thresholds, we use the subsampling method (200 out of 240 trials), which prove that the obtained thresholds are robust against outliers (Fig. 4e). The method reliably show different thresholds for the different stimulus frequencies, indicating, that the hearing ability of this animal is best in the range between 2-4 kHz (Fig. 4e), which corresponds to audiograms from the literature measured via behavioral paradigms [14].

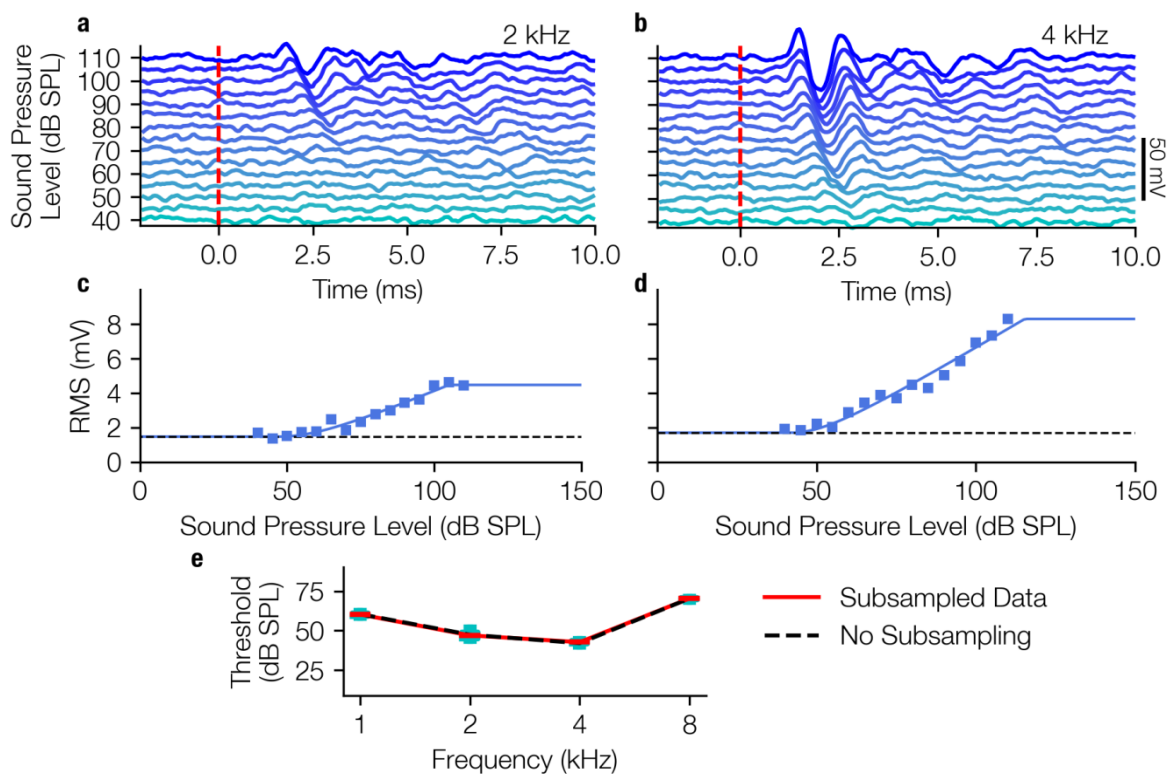

**Fig. 4 | ABR-data analysis.**
**a,b**, ABR waves (40-110 dB SPL, cyan to blue) of one animal (2kHz, 4kHz pure tone stimuli) averaged over 240 single trials (i.e. 120 double trials). **c,d**, Level response function (blue markers) approximated using the hard sigmoid fit (blue line). **e**, Audiogram with variances determined by subsampling (black dashed line: no subsampling, red line: medians, cyan boxes: quartiles, whisker 5%-95% percentiles, subsampling: N=200 out of 240 trials)**.**

## b) Cortical local field potential (LFP) data from different sensory modalities

The described fitting procedure can also be applied to other kinds of electrophysiological measures like cortical LFP data. Here we applied our new threshold estimation procedure to LFP recordings from the auditory and somatosensory cortex. Analogously to the processing of the ABR data, the RMS of the LFP data was used to estimate level response functions by fitting to the hard sigmoid function. For auditory stimulation pure-tone stimuli of 2 kHz were used, and for vibro-tactile stimulation 175 Hz stimuli were applied to the contralateral hind-limb of the animal via a Linear Resonant Actuator (cf. Methods).

For both auditory and somatosensory stimulation the stimulus response relationship can be described with the hard sigmoid function (Fig. 5b,e), yielding a threshold value through the knee criterion. Interestingly this method works equally well for sound stimuli sensed by the animals ear as well as for vibrational stimuli sensed via the animals paw. The analysis of LFP data recorded from both sensory modalities again demonstrates that our approach is robust against missing of supporting points near the threshold (cf. Fig. 5 b, e) and the number of applied measurement repetitions (Fig. 5c,f).

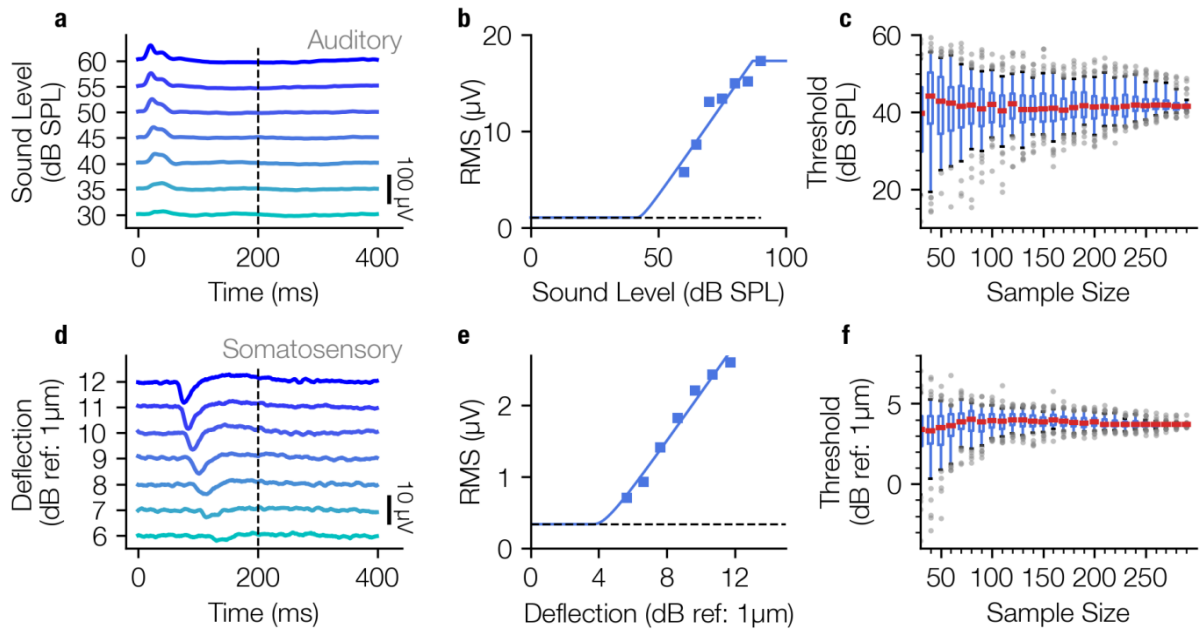

**Fig. 5 | Threshold determination using LFP data recorded in auditory and somatosensory cortex**
**a**,**b**,**c**, Determination of neuronal thresholds in the auditory cortex; **a**, LFP responses to 2 kHz tones of 200ms duration and varying sound pressure level (30-60 dB SPL). **b**, Level response function approximated using the hard sigmoid fit. **c**, Estimated threshold as a function of sample size (100 subsamples each). **d**, **e**, **f**: Analog analysis for LFP responses to vibro-tactile stimuli (200 ms, 175 Hz) of different amplitudes (6-12 dB) in the somatosensory cortex.

### c) Spiking data from auditory cortex

Finally, we apply our approach for threshold estimation to single neuron spiking data with just a few modifications, exemplarily shown for pure tone responses of auditory cortex neurons (Fig. 6). As spikes are detected by multi thresholding the raw signal, spike rates are not superposed by measurement noise. Instead, σ here is defined as the spontaneous activity, i.e. the spike rate in the absence of stimulation. As in contrast to LFP data no RMS values are calculated and spike rates can in first order be added, the fit function is the sum of hard sigmoid function and spontaneous activity (cf. equ. 2a, where $f_0(x)$ is the hard sigmoid function).

The measured spike rates show a clear dependency on the sound pressure level (Fig. 6a), which again can be described by a hard sigmoid function (cf. Fig. 6b). As already shown for the LFP data (Fig. 5), our novel fitting approach again needs a minimum of fitting parameters and is robust against the number of measurement repetitions (cf. Fig. 6c).

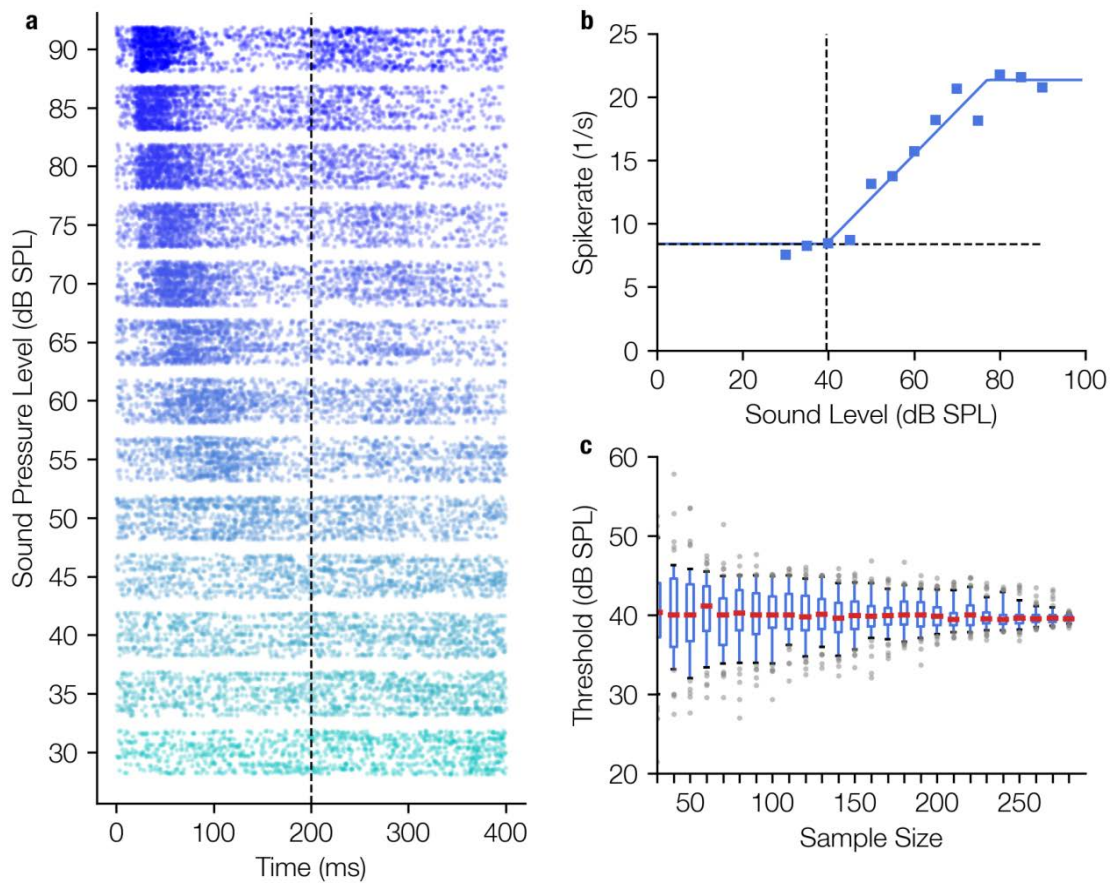

**Fig. 6 | Threshold determination using single unit spiking data recorded in the auditory cortex**
**a**, Dot raster diagram of the measured spikes (timestamps of spikes) for different stimulus intensities of a 2 kHz pure tone (30 - 90 dB SPL, onset 0ms, offset 200ms) and 300 measurement repetitions. The typical latency shift as a function of the applied sound pressure level can be observed. **b**, Hard sigmoid fit; For this fitting procedure the function $f_0(x) + \sigma$ is fitted in contrast to LFP recordings where the fit function is $\sqrt{f_0(x)^2 + \sigma^2}$. This difference arises from the fact that spontaneous spike rate and evoked spike rate can simply be added. Thus, the hard sigmoid added by the spontaneous activity term are a good description of the level response characteristics (b). **c**, The determined threshold shows no systematic dependency on the number of measurement repetitions.

# Discussion

In this article we present a novel and robust method for threshold estimation based on neurophysiological data. The robustness and objectiveness of the method is based on three main principles.

First, the sensory threshold is determined by the analysis of the complete amplitude response function and not only by analyzing the data near the threshold where the S/N ratio is worst. This advantage is achieved by applying a fitting procedure, where the measurement noise level σ is kept constant and is not treated as a free parameter. Thus, it is possible to estimate the shape of the amplitude response function by exclusively analyzing supporting points above the threshold. The validity of this procedure was systematically analyzed using artificial data (cf. Fig. 2) and verified by the analysis of different neurophysiological parameters. Taken together, the approach enables us to obtain an objective automated measurement of a threshold, eliminating the need of investigating "just-above" threshold responses within a noisy signal by visual threshold estimations by experimenters or clinical professionals [5,15-17].

The second principle is that the threshold estimation should not contain an explicit dependency on the measurement noise, i.e. the definition of the threshold as the minimum signal exceeding a certain fraction of the background noise amplitude (cf. 2σ criterion) leads to one major problem: These criteria depend systematically on the number of measurement repetitions, meaning that an increase of the S/N ratio of the measurement procedure leads to systematic decrease of the determined threshold. We have shown that the threshold as a function of the number of applied measurement repetitions does not asymptotically approach a constant value but diverges. To this end, we deleted any discrete dependency of the threshold criteria on the background noise.

The third major principle is the reduction of free selectable parameters. We have shown that fitting of an extended generalized logistic function could be used for threshold estimation, but needs one further free parameter defining the threshold. In contrast to that approach, we chose a hard sigmoid function with a clear defined knee (cf. Fig. 2e), which can be used to estimate the threshold without a pre-defined threshold criterion. Furthermore, the fit of a hard sigmoid function is more robust against missing points from the saturation range, thus the experiment does not need to cover the whole dynamic range of the sensory system to be investigated.

The method was evaluated on different sensory modalities (auditory and somatosensory), from different brain regions (cortex and brain stem), to demonstrate its wide and general applicability.

Though the method provides the possibility to estimate highly reproducible and plausible threshold estimates based on electrophysiological measurements it has to be considered that the determined threshold are only correlates of the "true" thresholds estimated using psychophysical methods.

Thus, thresholds based on electrophysiological measures led to frequency specific thresholds higher to what is known from the literature [14]. The reasons for this discrepancy may lie in the fact that the electrophysiological sum potentials are produced by several thousands of neurons, whereas a behavioral response can be evoked by the activity of a far smaller amount of firing neurons.

Additionally, the effects of anesthesia can have an effect on the determined thresholds, though it could be shown that the ketamine/xylazin anesthesia has no effect on ABR thresholds [18,19] in gerbils and rats but in mice [20].

Despite these limitations, a reproducible correlate for sensory thresholds is essential for scientific purposes as especially the objectiveness prevents the experimenter from systematic errors and reproducibility is a core concept of scientific studies.

# Online Methods

## Animals

Mongolian gerbils (*Meriones unguiculatus*) were housed in standard animal racks (Bio A.S. Vent Light, Ehret Labor- und Pharmatechnik, Emmendingen, Germany) in groups of 2 to 3 animals per cage with free access to water and food at 20 to 24°C room temperature under 12/12 h dark/light cycle. The use and care of animals was approved by the state of Bavaria (Regierungspräsidium Mittelfranken, Ansbach, Germany, No. 54-2532.1-02/13). All methods were performed in accordance with the relevant guidelines and regulations of NIH. A total of 11 male gerbils aged ten to twelve weeks purchased from Janvier Laboratories Inc. were used in this study.

## Data sources

### a) Generation of artificial test data sets

In order to evaluate our algorithm, artificial data sets were generated, to mimic far field ABR recordings (cf. Fig. 1). We assume that the response amplitude (root-mean-squared, RMS) as an answer to the stimulus intensity can be described by a sigmoid function (in this case a generalized logistic function):

$$f_0(x) = \frac{a}{1 + e^{-\frac{x-b}{c}}} \qquad \text{(equ. 1)}$$

Where *a* describes the maximal response, *b* defines the stimulus intensity where the response is exactly half the maximal response, and the factor *c* defines the slope of the sigmoidal shape. We assign the parameters with *a=10 mV, b=60 dB, c=11.89 dB* to obtain values in a realistic range and sample the function at 22 equally distributed points from -30 dB to 130 dB (Fig. 1A). A template amplitude response curve was used to generate the artificial field potential (Fig. 1A).

For each of those supporting points we generate a sinusoidal signal with a duration of 10 ms and an amplitude given by Equ. 1, to imitate the physiological response, e.g. the auditory brainstem activity (Fig. 1b). For each of these artificial responses, a measurement noise being Gaussian distributed noise (µ=0, σ=40mV, S/N=1/4) is created (Fig. 1c,d) and added to the pure response (Fig. 1d).

For every stimulus intensity *N*=200 such responses are created, to obtain *N* measurement repetitions (trials). The values of the parameters *a*, *b*, *c* as well as the noise amplitudes are chosen in accordance to real ABR measurements, however the sigmoid shape of the underlying stimulus response function is universal and thus the results can be applied to any kind of threshold determination tasks based on the interpretation of neural responses.

For each single trial, new noise is sampled, but with the same average background noise amplitude. This background noise reflects a combination of spontaneous neuronal activity and measurement noise. This physical noise being a superposition of several noise sources can be approximated with a Gaussian distributed noise of a certain amplitude.

From these (artificial) single trial responses, all *N* simulated repetitions for each stimulus intensity are averaged (mean simulated ABR responses, averaging 200 single trials as in Fig. 1d). The obtained averaged responses mimic the processes during a real measurement to provide data where the real underlying response amplitude function is known (cf. Fig.1e).

### b) ABR recordings

ABR were recorded using a custom made setup. Pure tone stimuli of different frequencies ranging from 1 to 8 kHz were generated by a custom-made Matlab program and presented at different, pseudorandomized intensities ranging from 40 to 110 dB SPL in 5 dB steps. Stimulation was free-field to one ear at a time via a speaker (Sinus Live NEO) corrected for its frequency transfer function to be flat within +/- 1 dB at a distance of approximately 3 cm from the animal's pinna while the contralateral ear was tamped with an ear plug. To compensate for speaker artifacts stimuli were presented in double trials consisting of two 6 ms stimuli (including 2 ms sine square rise and fall ramps) of the same amplitude but opposite phase, separated by 100 ms of silence. 120 to 500 double trials (*N*: Number of double trials) of each combination of intensity and frequency were presented pseudorandomly at an inter-stimulus interval of 500 ms.

For the measurements the Mongolian gerbils were kept under deep anesthesia. Anesthesia was induced by an initial dose of 0.3 ml of a ketamine-xylacin-mixture (mixture of ketamine hydrochloride: 96 mg/kg BW; xylacin hydrochloride: 4mg/kg BW; atropine sulfate: 1 mg/kg BW), and maintained by continuous application of that mixture at a rate of 0.2 to 0.3 ml/h. As has been demonstrated previously, such ketamin-xylazine anesthesia has only little effect on ABR signals compared to awake animals [19]. During measurements, animals were placed on a feedback-controlled heating pad at 37°C to maintain body temperature. Data were recorded using three silver electrodes positioned subcutaneously, one for grounding at the back of the animals, one reference electrode at the forehead, and the measuring electrode infra-auricular overlying the bulla contralateral to the stimulation side. The potential difference between the reference and measuring electrode was amplified by a low noise amplifier (JHM NeuroAmp 401, J. Helbig Messtechnik, Mainaschaff, Germany; amplification 10.000; bandpass filter 400 Hz to 2000 Hz and 50 Hz notch filter). Note that for further analysis the amplified signal was used, that is, amplitudes are given in mV whereas the actual neuronal signals were in µV-range. The output signal of the amplifier was digitalized and recorded by an analog-digital converter card (National Instruments Corporation, Austin, TX, USA) with a sampling rate of 20 kHz and synchronized with the stimulation via the trigger signal from the stimulation computer. Raw data of N double trials per sound level for one stimulus frequency were averaged. Finally, these averaged responses of the two single, phase inverted stimuli within one double trial were averaged to eliminate stimulus artifacts (Fig. 4a,b). From these averaged, artifact-corrected data the root mean square (RMS) values from 0 to 10 ms after stimulus onset were calculated to obtain a measure of response amplitude for each stimulus intensity presented (cf. Fig. 4c,d).

### c) Neuronal recordings in auditory and somatosensory cortex

For neuronal recordings in the primary sensory cortices a craniotomy was performed on Mongolian gerbils under deep ketamine-xylazine anesthesia as described above. During the complete surgery the animal was kept on a feedback-controlled heating pad at 37 °C to maintain body temperature, and the paw withdrawal reflex was checked periodically to ensure sufficient depth of anesthesia. A screw was fixed to the skull of the animal using instant glue and dental cement to provide a fixation during neuronal recordings. The neuronal recordings were performed directly after surgery. For recordings in auditory cortex [21,22], the animal was placed in an anechoic chamber on a heating pad, the head was fixed, and anesthesia was continued. Then a 16 electrode Pt-Ir array (Clunburry Scientific, 4x4 array, spacing 500 µm, Bloomfield Hills, MI 48304 USA) was inserted into the auditory cortex for LFP and single unit spike recording. For auditory threshold estimation 200 ms pure tone stimuli (2 kHz, 5ms ramp) of varying sound pressure level (30-90 dB SPL) were presented. The different sound pressure levels were presented pseudorandomly with 300 repetitions each.

Recordings in somatosensory cortex were performed as described above. For threshold measurements a linear resonant actuator specifically designed for haptic feedback application (LRA, C10-100 Precision Microdrives) was used to apply vibro-tactile stimuli to the hind limb of the animal (frequency: 175 Hz, range: ~0.9 µm- 15 µm).

## General Computation

All simulations and evaluation algorithm were run on a standard desktop PC and were written in Python using the Anaconda bundle. For mathematical operations the NumPy [23] and SciPy [24] library were used. The plots were created using the Matplotlib [25] library combined with the Pylustrator add-on [26] for plotting style editing.

## General fitting, threshold determination and subsampling of the data

In general, monotonous rate-intensity functions in neuronal systems (based on physiological data) follow a sigmoid function and, thus, are often described by a logistic function [27,28]. However, the superposition of measurement noise and stimulus induced neural signal can lead to slight changes in the shape of the stimulus response function. In the following, the effect of measurement noise on the curve shape as well as different threshold criteria based on the fit function are described in detail.

For all following analysis steps in addition to the fitting procedure a subsampling technique is applied. In all cases 100 different subsamples were generated. For each subsample *N-d* trials were randomly drawn, where $d > \sqrt{N}$ delete-d-jackkife criterion, cf. [29,30], without returning from the complete set of measured trials (*N*). The reduced set of measurement repetitions is used as base for the fitting procedure. This procedure is done for each stimulus intensity separately, so that each subsample contains N-d measured trials for each stimulus intensity.

The procedure is applied for two different reasons. First, the sample size is systematically altered to analyze the effect of different number of applied measurement repetitions on the determined threshold (in the following sample size is used synonymic to number of measurement repetitions), on the other hand for real neural data the method is used similarly to bootstrapping methods to estimate confidence intervals.

## Acknowledgements:
This work was supported by the Interdisciplinary Center for Clinical Research Erlangen (IZKF, project E15) and by the Deutsche Forschungsgemeinschaft (DFG-grant SCHU 1272/12-1).

## Author contributions
A.S., H.S. led the project. A.S., P.K., R.G., H.S. developed the method, to which C.M. made critical contributions. A.S., R.G. programmed the evaluation software. A.S., P.K. conducted the measurements. A.S., R.G., P.K., C.M., K.T., H.S. discussed the results. A.S., R.G., H.S. wrote the manuscript. C.M., K.T provided advice and article revisions. All authors approved the final version of the article.

## Conflict of Interest:
Authors report no conflict of interest.